\input jytex.tex   
\typesize=10pt
\magnification=1200
\baselineskip17truept
\footnotenumstyle{arabic}
\hsize=6truein\vsize=8.5truein
\sectionnumstyle{blank}
\chapternumstyle{blank}
\chapternum=1
\sectionnum=1
\pagenum=0

\def\begintitle{\pagenumstyle{blank}\parindent=0pt\begin{narrow}[0.4in]}
\def\endtitle{\end{narrow}\newpage\pagenumstyle{arabic}}


\def\beginexercise{\vskip 20truept\parindent=0pt\begin{narrow}[10
truept]}
\def\endexercise{\vskip 10truept\end{narrow}}


\def\eql#1{\eqno\eqnlabel{#1}}
\def\ref{\reference}
\def\peq{\puteqn}
\def\pref{\putref}

\def\mgn{\marginnote}
\def\bex{\begin{exercise}}
\def\eex{\end{exercise}}


\font\open=msbm10 
\def\mbox#1{{\leavevmode\hbox{#1}}}

\def\hspace#1{{\phantom{\mbox#1}}}
\def\oR{\mbox{\open\char82}}

\def\oZ{\mbox{\open\char90}}

\def\al{\alpha}

\def\ga{\gamma}
\def\de{\delta}
\def\Ga{\Gamma}

\def\la{\lambda}
\def\La{\Lambda}

\def\ze{\zeta}

\def\De{\Delta}

\def\Det{{\rm Det\,}}
\def\Real{{\rm Re\,}}

\def\zf{$\zeta$--function}
\def\zfs{$\zeta$--functions}


\def\frac#1/#2{\leavevmode\kern.1em
\raise.5ex\hbox{\the\scriptfont0 #1}\kern-.1em/\kern-.15em
\lower.25ex\hbox{\the\scriptfont0 #2}}
\def\sfrac#1/#2{\leavevmode\kern.1em
\raise.5ex\hbox{\the\scriptscriptfont0 #1}\kern-.1em/\kern-.15em
\lower.25ex\hbox{\the\scriptscriptfont0 #2}}

\def\gtorder{\mathrel{\raise.3ex\hbox{$>$}\mkern-14mu
             \lower0.6ex\hbox{$\sim$}}}
\def\ltorder{\mathrel{\raise.3ex\hbox{$<$}\mkern-14mu
             \lower0.6ex\hbox{$\sim$}}}

\def\semidirprod{\rlap{\ss C}\raise1pt\hbox{$\mkern.75mu\times$}}
\def\for{\lower6pt\hbox{$\Big|$}}
\def\fish{\kern-.25em{\phantom{abcde}\over \phantom{abcde}}\kern-.25em}


\def\boxit#1{\vbox{\hrule\hbox{\vrule\kern3pt
        \vbox{\kern3pt#1\kern3pt}\kern3pt\vrule}\hrule}}
\def\dalemb#1#2{{\vbox{\hrule height .#2pt
        \hbox{\vrule width.#2pt height#1pt \kern#1pt
                \vrule width.#2pt}
        \hrule height.#2pt}}}

\def\frac#1#2{{{#1}\over{#2}}}

\def\comb#1#2{{\left(#1\atop#2\right)}}

\def\eg{{\it e.g. }}
\def\ie{{\it i.e. }}
\def\cf{{\it cf }}
\def\pa{\partial}


\def\sgap{\vskip 15truept}

\def\sumdash#1{{\mathop{{\sum}'}_{#1}}}

\def\sumstar#1{{\mathop{{\sum}^*_{#1}}}}
\def\prodstar#1{{\mathop{{\prod}^*_{#1}}}}

\def\3j#1#2#3#4#5#6{\left\lgroup\matrix{#1&#2&#3\cr#4&#5&#6\cr}
\right\rgroup}

\def\man{{\cal M}}

\def\m?{\mgn{?}}

\def\pa{\partial}

\def\beq{\begin{eqnarray}}
\def\eeq{\end{eqnarray}}


\def\aop#1#2#3{{\it Ann. Phys.} {\bf {#1}} ({#2}) #3}

\def\cmp#1#2#3{{\it Comm. Math. Phys.} {\bf {#1}} ({#2}) #3}
\def\cqg#1#2#3{{\it Class. Quant. Grav.} {\bf {#1}} ({#2}) #3}

\def\jmp#1#2#3{{\it J. Math. Phys.} {\bf {#1}} ({#2}) #3}
\def\jpa#1#2#3{{\it J. Phys.} {\bf A{#1}} ({#2}) #3}

\def\np#1#2#3{{\it Nucl. Phys.} {\bf B{#1}} ({#2}) #3}
\def\pl#1#2#3{{\it Phys. Lett.} {\bf {#1}} ({#2}) #3}

\def\pr#1#2#3{{\it Phys. Rev.} {\bf {#1}} ({#2}) #3}
\def\prA#1#2#3{{\it Phys. Rev.} {\bf A{#1}} ({#2}) #3}

\def\prD#1#2#3{{\it Phys. Rev.} {\bf D{#1}} ({#2}) #3}

\def\prs#1#2#3{{\it Proc. Roy. Soc.} {\bf A{#1}} ({#2}) #3}
\def\pcps#1#2#3{{\it Proc. Camb. Phil. Soc.} {\bf{#1}} ({#2}) #3}

\def\am#1#2#3{{\it Acta Mathematica} {\bf {#1}} ({#2}) #3}
\def\aim#1#2#3{{\it Adv. in Math.} {\bf {#1}} ({#2}) #3}

\def\aom#1#2#3{{\it Ann. of Math.} {\bf {#1}} ({#2}) #3}

\def\cpde#1#2#3{{\it Comm. Partial Diff. Equns.} {\bf {#1}} ({#2}) #3}

\def\dmj#1#2#3{{\it Duke Math. J.} {\bf {#1}} ({#2}) #3}

\def\jdg#1#2#3{{\it J. Diff. Geom.} {\bf {#1}} ({#2}) #3}
\def\jfa#1#2#3{{\it J. Func. Anal.} {\bf {#1}} ({#2}) #3}

\def\ma#1#2#3{{\it Math. Ann.} {\bf {#1}} ({#2}) #3}
\def\mom#1#2#3{{\it Messenger of Math.} {\bf {#1}} ({#2}) #3}
\def\mz#1#2#3{{\it Math. Zeit.} {\bf {#1}} ({#2}) #3}
\def\ojm#1#2#3{{\it Osaka J.Math.} {\bf {#1}} ({#2}) #3}
\def\pams#1#2#3{{\it Proc. Am. Math. Soc.} {\bf {#1}} ({#2}) #3}

\def\pja#1#2#3{{\it Proc. Jap. Acad.} {\bf {A#1}} ({#2}) #3}

\def\qjm#1#2#3{{\it Quart. J. Math.} {\bf {#1}} ({#2}) #3}

\def\rmjm#1#2#3{{\it Rocky Mountain J. Math.} {\bf {#1}} ({#2}) #3}

\def\top#1#2#3{{\it Topology} {\bf {#1}} ({#2}) #3}
\def\tams#1#2#3{{\it Trans.Am.Math.Soc.} {\bf {#1}} ({#2}) #3}

\begin{title}
\vglue 1truein
\vskip15truept
 \centertext {\Bigfonts \bf The Barnes $\zeta$--function, sphere determinants
and}
\vskip10truept
\centertext{\Bigfonts \bf Glaisher-Kinkelin-Bendersky constants}
\vskip 20truept
\centertext{J.S.Dowker\footnote{ dowker@a35.ph.man.ac.uk}}
\vskip 7truept \centertext{\it
Department of Theoretical Physics, } \centertext{\it The University of
 Manchester,} \centertext{\it Manchester, England}
 \vskip 15truept
\centertext{Klaus Kirsten\footnote{ klaus\_kirsten@baylor.edu}}
\vskip 7truept \centertext{\it Department of Mathematics,}
\centertext{\it Baylor University,}
\centertext{\it Waco, TX 76798, USA}
 \vskip 15truept \centertext{Abstract} \vskip8truept
\begin{narrow}
Summations and relations involving the Hurwitz and Riemann \zfs\ are
extended first to Barnes \zfs\ and then to \zfs\ of general type. The
analysis is motivated by the evaluation of determinants on spheres which
are treated both by a direct expansion method and by regularised sums.
Comments on existing calculations are made. It is suggested that the
combination $\ze'(-n)+H_n\ze(-n)$, where $H_n$ is a harmonic number, should
be taken as more relevant than just $\ze'(-n)$. This leads to a
Kaluza--Klein technique, providing a determinant interpretation of the
Glaisher--Kinkelin--Bendersky constants which are then generalised to
arbitrary \zfs. This technique allows an improved treatment of sphere
determinants.
\end{narrow}
\vskip 5truept
\vskip 60truept
\vfil
\end{title}
\pagenum=0
\section{\bf1. Introduction}
The Barnes \zf, [\pref{Barnesa}], has enjoyed only sporadic mathematical
interest since it was first introduced in 1903, [\pref{Vign, Shint, Kuro}].
It generalises the Hurwitz \zf, has number theory applications and involves
a natural generalisation of the Euler $\Ga$--function, examples of which had
already appeared, [\pref{kink,holder,alex}] and were developed more
directly by Bendersky [\pref{bend}].

Some of these works discuss regularised products,
[\pref{QHS,JandL,Illies,Manin}], and the related Laplacian determinants,
[\pref{Vardi,Voros,Sarnak}], on various spaces from varying points of view.
Explicit results are usually obtained only when the eigenvalue spectrum is
known or sufficiently powerful geometrical or analytical information is
available.

As might be expected, particularly detailed investigations have been made
in the case of spheres. Explicit $d$-sphere determinant
expressions are given, and
plotted, in [\pref{Dow3}], obtained by a direct method, ancillary to a
discussion of the orbifold quotient, S$^d/\Ga$. Choi [\pref{Choi}] analyses
the $3$-sphere using a factorisation technique founded on the method in
Voros' basic paper [\pref{Voros}] while Choi and Srivastava [\pref{ChandS}]
improve this calculation. Quine and Choi [\pref{QandC}], with a systematic
method, give explicit forms for the full $d$-sphere. Kumagai
[\pref{kumagai}] attempts to give a corrected version of Vardi's analysis,
[\pref{Vardi}],
for the $d$-sphere but does not calculate beyond the 4-sphere.\footnote
{ As noted in [\pref{Dow3}], it is clear there is some error in Vardi's
manipulations. This can be traced to two simple oversights on p.504 of
[\pref{Vardi}]. The
first is in the proof that there are no $\log k$ terms. The upper limit on
the $k$ sum cannot be changed from $n-1$ to $n-2$ because there is a
contribution when $d=0$. The second slip is an incorrect interchange of the
$d$ and $r$ summations. Vardi produces a general formula for the
$d$--sphere determinant in terms of the derivatives of the Riemann \zf\ at
negative integers, \cf [\pref{Dow3,QandC}], which he then converts into
multiple $\Ga$--functions, $\Ga_n(1/2)$.} Further remarks are made later.

During the course of these investigations, identities and relations appear
which we would like to systematise in a certain way and enlarge upon. Some
considerable amount of work in this area has appeared in the mathematical
literature in the past few years and we wish to draw attention to relevant
work by physicists which is often overlooked and which might be useful.

Applications of the Barnes function in physics embrace the Casimir effect
around cosmic strings [\pref{Dow4}], on orbifolded spheres, [\pref{ChandD,
Dow1}], integrable field theories [\pref{Smirnov,Ruijsenaars,Luk, FandZ}],
and statistical mechanics, [\pref{JandM}]. The Barnes \zf\ also arises for
the higher-dimensional harmonic oscillator and is useful in connection with
Bose-Einstein condensation and trapping [\pref{KandT,HKK}].

The structure of this paper is as follows. The Barnes \zf\ is introduced
via scalar \zfs\ on spheres. The determinant is next looked at, which
focuses attention on certain constructs and relations. These further
motivate an investigation of the {\it general} \zf, $\sum \la^{-s}$, and we
then link up with the regularised product approach giving a Kaluza-Klein
interpretation of generalised Glaisher-Kinkelin-Bendersky constants. The
paper proceeds as a series of generalisations.

Determinants appear frequently in field and string theory. No attempt will
be made to justify their computation nor to detail their general history.
For the 2-sphere the first computations were by Hortacsu {\it et al}
[\pref{Hortacsu}] and by Weisberger [\pref{Weisbergera,Weisbergerb}]. The
topic of \zfs\ on spheres, and symmetric spaces in general, also has a long
record. (See Camporesi, [{\pref{Camporesi}], for a useful survey and
results.)
\section{\bf2. The sphere \zf\ and the Barnes \zf.}

The general definition of the Barnes \zf\ is,
  $$\eqalign{ \zeta_d(s,a|{\bf {d}})=&{i\Gamma(1-s)\over2\pi}\int_L
  d\tau {\exp(-a\tau)
  (-\tau)^{s-1}\over\prod_{i=1}^d\big(1-\exp(-d_i\tau)\big)}\cr
  =&\sum_{{\bf {m}}={\bf 0}}^\infty{1\over(a+{\bf {m.d}})^s},\qquad
  \Real\, s>d\,,} \eql{barn}
  $$
where we refer to the components, $d_i$, of the $d$-vector, ${\bf d}$, as
the {\it degrees} or {\it quasi-periods}. For simplicity, we assume that
the $d_i$ are real and positive. If $a$ is zero, the origin, ${\bf m=0}$,
is to be excluded. The contour, $L$, is the standard Hankel one.

On the $d$-sphere, consider the Laplacian--type operator,
  $$
  -\De+\xi R-\al^2\,,
  \eql{op}$$
where $\xi=(d-1)/4d$ and $\al$ is a parameter introduced for convenience of
discussing several specific cases at once. The value $\al=0$ corresponds to
conformal coupling in $d+1$ dimensions, $\al=1/2$ to conformal coupling in
$d$-dimensions and $\al=(d-1)/2$ to minimal coupling \ie just the operator
$-\De$. In the first case the eigenvalues are perfect squares,
  $$
  {1\over4}(m+d-2)^2,\quad m =1,3,\ldots\,.
  \eql{eig}$$

It is best to think of the mode set on the {\it full} sphere as the union
of Dirichlet and Neumann mode sets on a {\it hemisphere}, despite the
apparent extra complication. The reason is that these individual \zfs\ are
Barnes functions, with all degrees equal to unity, ${\bf d=1}$, as shown in
[\pref{ChandD}].  Specifically, for $\al=0$,
  $$
  \zeta_{{}_N}(s)=\zeta_d\left(2s,(d-1)/2\mid {\bf d}\right), $$
  $$
  \zeta_{{}_D}(s)=\zeta_d\left(2s,{\textstyle\sum}\,d_i-(d-1)/2\mid
  {\bf d} \right),
  $$
where the parameters in the \zfs\ have been left general because, although
for the full--sphere, and hemisphere, we need only unit degrees, we wish to
retain the general case for as long as possible as it applies for the other
orbifold factorings of the sphere\footnote { In this case the $d_i$ are the
integer {\it degrees} associated with the polytope symmetry group, $\Ga$.
For the simplest (cyclic) case, $\oZ_q$, $d_1=q$ with the rest unity.}.

If the sum over the vector ${\bf m}$ in (\peq{barn}) is performed as
far as possible, it is easy to regain the standard eigenvalues,
(\peq{eig}), together with the usual degeneracies.

For the general operator (\peq{op}) the \zf\ to define is clearly,
  $$
\ze(s,a,\al;{\bf d})=\sum_{\bf m=0}^\infty {1\over\big((a+{\bf m.d})^2-
\al^2\big)^s}\,,
  \eql{zeta1}$$ and if ${\bf d=1}$, giving the hemisphere, we have,
  $$
\ze_{HS}(s)=\sum_{m=0}^\infty\comb{m+d-1}{d-1}{1\over\big((a+m)^2-
\al^2)\big)^s}.
  \eql{zeta2}$$
From this, it is easy to confirm that the sum of the Dirichlet and Neumann
\zfs\ on the hemisphere does equal the full--sphere \zf\ [\pref{Dow3}]. A
technical point is that for minimal coupling and Neumann conditions, the
origin, ${\bf m=0}$, is to be omitted from the sum.

There are at least three approaches to the continuation of the \zf\
(\peq{zeta2}). One is to use a summation formulae, such as Plana's (see
[\pref{Dow13}] for this specific case). The second is that used by
Minakshisundaram and later by Candelas and Weinberg [\pref{CandW}] in a
treatment of sphere \zfs\ and employs a Bessel function identity to perform
the summation over $m$. The third method involves an expansion in the
associated parameter, $\al^2$, and, after some manipulation with the
binomial coefficient, the problem is thrown onto the continuation of a
series of Hurwitz \zfs, a standard matter \footnote{ For $\al=0$, \ie the
Barnes \zf, Barnes [\pref{Barnesb}] gives the reduction to a finite sum of
Hurwitz \zfs\ and remarks that it {\it could} be made fundamental for the
theory when ${\bf d=1}$. This reduction has been rediscovered in many, more
recent discussions of the multiple $\Ga$-- function, restricted, as they are,
to just this case. In particular, on p.432, Barnes gives an expression
for $\log\Ga_n$ essentially equivalent to formulae of Vardi [\pref{Vardi}]
and Kanemitsu {\it et al}, [\pref{KKY}]. A similar reduction is also
possible for rational degrees.}.

The first approach is not practicable for general degrees, ${\bf d\ne1}$,
(\peq{zeta1}), since there are several genuine summation variables. The
second method yields the Bessel function expression
  $$
  \ze(s,a,\al\mid{\bf d})={\sqrt\pi\over\Ga(s)}\int_0^\infty {d\tau
  \exp(-a\tau)\over\prod_{i=1}^d(1-\exp(-d_i\tau))}\bigg({\tau\over2\al}\bigg)
  ^{s-1/2}\,I_{s-1/2}(\al\tau)\,, \eql{bess}
  $$
which could be continued to give the analogue of (\peq{barn}) or treated as
in Candelas and Weinberg [\pref{CandW}] and Chodos and Myers
[\pref{Chodos1}] to enable values at negative $s$ to be computed.

This is a practical method of obtaining a continuation of the sphere \zf\
but as our main interest is really with the Barnes function we apply the
third procedure and expand in $\al^2$. One then encounters the continuation
of an infinite series of Barnes \zfs, suitable only for special values of
$s$ but sufficient for our purposes.
\section{\bf3. Sphere determinants.}

The first step is a standard expansion in powers of $\al^2$, (\cf
[\pref{Wilton}]),
 $$
\ze(s,a,\al\mid{\bf d})=\sum_{r=0}^\infty\al^{2r}{s(s+1) \ldots(s+r-1)\over
r!}\,\ze_d(2s+2r,a\mid {\bf d}),
  \eql{expan}$$
which allows certain information to be extracted, but which does not
constitute a complete continuation. The principle being applied here is
expansion in terms of `known' \zfs\ and the continuation of the Barnes \zf\
will be assumed in principle to be already achieved by Barnes with the
numerical computation of any particular case looked upon as a soluble
technical challenge.

Barnes has given the values of $\ze_d(s,a\mid{\bf d})$ for $s$ a
nonpositive integer, and also the residues and remainders at the poles
$s=1,\ldots,d$, in terms of generalised Bernoulli functions
rapidly computed by recursion. Hence, for example, one can easily find the
values $ \ze(-n,a,\al\mid{\bf d})$ for $n=0,-1,\ldots$, [\pref{Dow1}].

The derivative at 0 also follows, but not so directly. Firstly from
(\peq{expan})
  $$\eqalign{
\ze'(0,a,\al\mid{\bf d})=2\ze_d'(0,a\mid{\bf d})+\sum_{r=1}^u{\al^{2r}
\over r}&\left(R_{2r}(d)+ {1\over2}N_{2r}(d)\sum_{k=1}^{r-1}{1\over
k}\right)\cr +&\sum_{r=u+1}^\infty{\al^{2r}\over r}\ze_d(2r,a\mid{\bf d}).}
\eql{spderiv}
  $$
The notation is that $N_r(d)$ is the residue and $R_r(d)$ the remainder
defined by the behaviour at the known Barnes poles\footnote{ It may happen
that a residue vanishes.}
  $$
\ze_d(s+r,a\mid{\bf d})\rightarrow {N_r(d)\over s}+R_r(d)
\quad{\rm as}\,\,s\rightarrow 0,
  \eql{pole1}
  $$
where $1\le r\le d$ and $u=d/2$ if $d$ is even and $u=(d-1)/2$ if $d$ is
odd.

The problem is the evaluation of the infinite series on the right.
\section{\bf4. \zf\ sums.}
Expression (\peq{spderiv}) concentrates our attention on the sum of
Barnes \zfs,
  $$
\sum_{r=u+1}^\infty{\al^{2r}\over r}\,\ze_d(2r,a\mid{\bf d})\,, \eql{sum1}
  $$
which can be obtained from
  $$
\sum_{r=d+1}^\infty{(-\al)^r\over r}\,\ze_d(r,a\mid{\bf d})\,,
  \eql{sum2}$$
by averaging over $\pm\al$.

For the Hurwitz \zf\ ($d=1$) this sum is standard (\eg [\pref{WandW}]
p.276) \footnote{ More complicated sums involving the Hurwitz \zf\ have
been extensively investigated by Srivastava and others, \eg [\pref{Sriv}].
Several methods have been adopted and can give different `final'
expressions leading to identities.} but the analysis extends to the more
general case as shown by Barnes who has, ([\pref{Barnesa}] p.424 eq(1)),
(\cf [\pref{Dow12}]),
  $$
  \sum_{r=d+1}^\infty{(-\al)^r\over r}\,\ze_d(r,a\mid{\bf d})= \log
  {\Ga_d(a+\al)\over\Ga_d(a)}-\sum_{r=1}^d{\al^r\over r!}\,
  \psi_d^{(r)}(a)\,.  \eql{sum3}
  $$
The sum on the left is over those $\psi$ functions defined by the basic
summation formula while that on the right contains $\psi$'s that have to be
regularised.

In this paper we maintain Barnes' original notation so that the 
multiple $\Ga$-- and $\psi$--functions are defined by
  $$
\ze'_d(0,a\mid{\bf d})=\log {\Ga_d(a)\over\rho_d({\bf d})}\,,\quad
\psi_d^{(q)}(a)={\pa^q\over \pa a^q}\log\Ga_d(a)\,,
  \eql{defs}$$
where  the $(d+1)$th $\Ga$--modular form, $\rho_d$, is given by
  $$
  \lim_{a\to0}\big[\ze'_d(0,a\mid{\bf d})+\log a\big]=-\log \rho_d({\bf d})\,.
  \eql{gamod}$$

Before generalising (\peq{sum3}), we derive it by the method of [\pref{Dow1}]
for a reason explained later.

The integral representation of the Barnes \zf\ allows the left-hand
sum in (\peq{sum3}) to be written, using an intermediate
regularisation,
  $$\eqalign{
\sum_{r=d+1}&{(-\al)^r\over r\,\Ga(r)}\int_0^\infty d\tau\tau^{r-1}
K(\tau)\cr =&\lim_{s\to0} \int_0^\infty d\tau\,
\left(\exp(-\al\tau)-\sum_{r=0}^d{(-\al\tau)^{r} \over
r!}\right)\tau^{s-1}K(\tau)\cr
=&\lim_{s\to0}{\bigg(\Ga(s)\ze_d(s,a+\al\mid\bf
d})-\sum_{r=0}^d{(-\al)^r\over r!} \Ga(s+r)\ze_d(s+r,a\mid{\bf
d})\bigg)\,,}
  \eql{reg}$$
where the `heat-kernel' $K(\tau)$ is defined by
  $$
  K(\tau)={\exp(-a\tau)\over\prod_{i=1}^d\big(1-\exp(-d_i\tau)\big)}\,.
  $$

Since the total quantity in (\peq{reg}) is finite, the individual pole
terms that arise in the $s\to0$ limit must cancel yielding the identity
between generalised Bernoulli polynomials,
  $$
\ze_d(0,a+\al\mid{\bf d})-\ze_d(0,a\mid{\bf d})
=\sum_{r=1}^d{(-\al)^{r}\over r}N_{r}(d).  \eql{iden1}
  $$

The finite remainder is the required result and equals,
  $$\eqalign{
\ze_d'(0,a+\al\mid{\bf d})&-\ze_d'(0,a\mid{\bf d}) -\sum_{r=1}^d
{(-\al)^{r}\over r}R_{r}(d)\cr &-\ga\big(\ze_d(0,a+\al\mid{\bf
d})-\ze_d(0,a\mid{\bf d})\big)- \sum_{r=1}^d {(-\al)^{r}\over
r}\psi(r)N_{r}(d)\,,}
  $$
which, in view of (\peq{iden1}), yields
  $$\eqalign{ & \sum_{r=d+1}^\infty{(-\al)^r\over
r}\,\ze_d(r,a\mid{\bf d})\cr &=\ze_d'(0,a+\al\mid{\bf
d})-\ze_d'(0,a\mid{\bf d})- \sum_{r=1}^d {(-\al)^{r}\over
r}R_{r}(d)-\sum_{r=1}^d {(-\al)^{r}\over r}
\big(\psi(r)+\ga\big)N_{r}(d)\cr &= \log {\Ga_d(a+\al)\over\Ga_d(a)}-
\sum_{r=1}^d {(-\al)^{r}\over r}\big(R_{r}(d)+
H_{r-1}\,N_{r}(d)\big)\,,\cr} \eql{sum4}
  $$
where $H_r$ is the harmonic number, $H_r=\sum_{k=1}^r\,1/k$, with $H_0=0$.

To compare with (\peq{sum3}) the form of the remainder $R_r(d)$ given
in Barnes is
 $$
 R_r(d)=(-1)^r\bigg({1\over(r-1)!}\psi_d^{(r)}(a)-N_r(d)H_{r-1}\bigg)\,,
 $$
 and we see that the two expressions for the sum, (\peq{sum3}) and
(\peq{sum4}), agree.

We will return to the sphere derivative (\peq{spderiv}) in section 8,
but proceed to generalise the sum (\peq{sum3}) by differentiating with
respect to $\al$, multiplying by $\al^\la$ and integrating back to
get,
  $$\eqalign{ \sum_{r=d+1}^\infty{(-1)^r\al^{r+\la}\over r+\la}\,
  \ze_d(r,a\mid{\bf d})= \int_0^\al dw\,&
  w^\la\big(\psi^{(1)}_d(a+w)-\psi^{(1)}_d(a)\big)\cr
  &-\sum_{r=2}^d{\al^{r+\la}\over (r+\la)(r-1)!}\, \psi_d^{(r)}(a)\,.}
  \eql{sum5}$$

This formula generalises one given essentially by Nash and O'Connor
[\pref{NandOC}] for the Hurwitz case, $d=1$, when it follows after simple
geometric summation. (See also Choi and Nash [\pref{CandN}].)

Clearly one could continue playing the same game and derive similar
summations but further progress can be made with (\peq{sum5}) when
$\la=n=0,1,2,\ldots$. We concentrate on the integral on the right-hand side
which we write in the form
  $$
\int_0^\al dw\, w^n\psi^{(1)}_d(a+w) =
\al^n\log\Ga_d(a+\al)-\de_{n0}\log\Ga_d (a) -n\int_0^\al dw\,w^{n-1}\log
\Ga_d(a+w),
  \eql{paint}$$
with the understanding that the final term vanishes when $n=0$.
\section{\bf5. Moments of $\log\Ga$ and $\psi$.}
As a preliminary, we discuss integrals of the type
  $$ \int_0^\al dw\, w^{n-1}\,\log\Ga_d(w)\,,\quad\int_0^\al dw\,
  w^n\psi^{(1)}_d(w)\,, \eql{ints}$$ which can be treated by the
  method in [\pref{DandKi}] since the algebraic technique
  applies equally well to the Barnes \zf, or indeed to any `modified'
  \zf\ of the form,\footnote{ This occurs in many places and $w$ could
  be thought of as a (mass)$^2$ or as a Laplace transform/resolvent
  variable.}
  $$
  \ze(s,w)=\sum_m{1\over(\la_m+w)^s}\,, \eql{genzet}$$ as it relies
  solely on iteration of the basic relation
  $$
  {\pa\ze(s,w)\over\pa w}=-s\ze(s+1,w)\,.
  \eql{reln2}$$

The formulae in [\pref{DandKi}] give indefinite integrals. We here choose
the definite forms, (\peq{ints}) with a lower limit of zero. In the Barnes
case, the only difference
is that $\log\sqrt{2\pi}$ becomes $\log\rho_d$ but this can be avoided
formally by extending the summation in [\pref{DandKi}] eq.(109) down to
$l=0$ and then making some algebraic transformations to arrive at the
result,
  $$\eqalign{ {1\over n!}\int_0^\al dw\,
  w^n\psi^{(1)}_d(w)=&-\sum_{i=1}^d{(-1)^n\over n!}
  \big(\ze'_{d+1-i}(-n,d_i\mid{\bf d}_i)
  +H_n\,\ze_{d+1-i}(-n,d_i\mid{\bf d}_i)\big)\cr
  &+\sum_{l=0}^n{\al^{n-l}\over(n-l)!}{(-1)^l\over l!} \big(\ze'_d(-l,
\al\mid{\bf
  d})+H_l\,\ze_d(-l,\al\mid{\bf d})\big)\,.}
  \eql{psimom1}$$

The first term on the right-hand side of (\peq{psimom1}) comes from the
lower limit, $w=0$, which we have treated by iterating the basic recursion,
  $$
  \ze_d(s,w\mid{\bf d})=\ze_d(s,w+d_*\mid{\bf
  d})+\ze_{d-1}(s,w\mid{\bf d}_*)\,,
  \eql{iter}$$
where $d_*$ is any degree and ${\bf d_*}$ is the $(d-1)$-vector obtained by
omitting the $d_*$ component from ${\bf d}$. Equation (\peq{iter}) has been
iterated on $d$ down to $\ze_0(s,w)=1/w^s$. For convenience we have chosen
the $d_*$ to be $d_1,d_2,\ldots,d_d$ taken in turn. The notation in
(\peq{psimom1}) then is that the vector ${\bf d}_i$ has components
$(d_{i},\ldots,d_d)$, \eg ${\bf d}_1={\bf d}$ and ${\bf d}_d$ is just the
single number $d_d$. Finally we let $w$ tend to zero. Incidentally,
(\peq{iter}) reveals why $d_*$ is a {\it quasi}--period.\footnote{ For the
standard Hurwitz \zf, the single quasi-period is 1, while for the general
case, (\peq{genzet}), there are none.}

Adamchik derives the result (\peq{psimom1}) for the simpler Hurwitz
function ($d=1$) from a series expression for $\log\Ga(1+x)$. Espinosa and
Moll [\pref{EandM}] use recursion to arrive at the Hurwitz results.
\section{\bf6. The general \zf.}
It is clear from its structure that (\peq{sum3}) is the regularised form of
an eigenvalue sum and its derivation can be paralleled formally for the
general \zf, (\peq{genzet}), the precise form of the heat-kernel in
(\peq{reg}) not being required.\footnote{ The identity, (\peq{iden1}),
becomes a standard one between heat-kernel coefficients.} The result is
exactly equation (\peq{sum4}), rewritten in a slightly different
notation,\footnote{ An alternative derivation using regularised sums is
contained in the next section.}
  $$
\sum_{r={[\mu]}+1}^\infty{(-\al)^r\over r}\,\ze(r,w)= \log
{\Ga(w+\al)\over\Ga(w)}- \sum_{r=1}^{[\mu]} {(-\al)^{r}\over
r}\big(FP\ze(r,w)+ H_{r-1}N_{r}(w)\big)\,,
  \eql{sum6}$$
where $\mu$ is the order of the infinite set $\la_m$, which could be any
sequence of numbers, \eg [\pref{Voros}], but which most commonly arises as
the spectrum of a self-adjoint operator, typically the Laplacian on a
Riemannian manifold when $\mu=d/2$, where $d$ is the dimension of the
manifold. However, one should not limit the meaning of the $\la_m$ to this
case. As they have arisen here, they could be the quantities ${\bf m.d}$ in
(\peq{zeta1}) and might be interpreted as the spectrum of a
pseudo-differential operator, such as the square root, $\De^{1/2}$, or
something similar. The ${\bf m.d}$ can be realised as the eigenvalues of
the pseudo--operator $-i{\bf d.\,|\nabla|}$ on the $d$--torus, the degrees,
$d_i$, being the inverse radii and $\mu$ being the dimension, $d$. In
symmetrical cases, for example when several degrees coincide, a partially
spherical realisation is possible, in accordance with section 2.

The pole residues are given in terms of the coefficients in the
short--time expansion of the heat-kernel,
$\sum\exp\big(-(\la_m+w)\tau\big)$ (including the mass-squared term,
$w$) in a standard way,
  $$
  N_r(w)={C_{\mu-r}(w)\over(r-1)!}\,.
  \eql{resi}$$
The finite part, $FP\ze(r,w)$, is just another symbol for the remainder,
$R_r$, at the {\it possible} $s=r$ pole,
 $$
\ze(s+r,w)\rightarrow {N_r(w)\over s}+R_r(w)
\quad{\rm as}\,\,s\rightarrow 0\,.
  \eql{pole2}$$

For a given sequence, $\la_m$, there may be no relevant poles at
$s=r\in\oZ$ so that $FP\ze(r,w)=\ze(r,w)$ and there would be no
need to separate the summations as in (\peq{sum6}).

The $\Ga$-functions in (\peq{sum6}) are defined in the usual manner,
\cf (\peq{gamod}), by
  $$
  \ze'(0,w)=\log {\Ga(w)\over\rho},\quad \log\rho=-\widetilde\ze'(0)\,,
  \eql{defs2}$$
with the `massless' \zf, denoted by a tilde,
  $$\widetilde\ze(s)=\sumdash{}{1\over\la^s_m}\,,
  $$
any zero modes being omitted. With this convention, one could set
$\widetilde\ze(s)=\ze(s,0)$.

If $\psi$--functions are also defined in the usual fashion,
 $$
  \psi^{(q)}(w)={\pa^q\over\pa w^ q}\,\log\Ga(w)\,, \eql{defs3}$$ so
  that for $q>\mu$,
  $$
  \psi^{(q)}(w)=(-1)^q(q-1)!\,\ze(q,w),
  $$
the general formula, (\peq{sum6}), looks {\it exactly} like the Barnes
formula, (\peq{sum3}),
  $$
  \sum_{r=[\mu]+1}^\infty{(-\al)^r\over r}\,\ze(r,w)= \log
  {\Ga(w+\al)\over\Ga(w)}-\sum_{r=1}^{[\mu]}{\al^r\over r!}\,
  \psi^{(r)}(w)\,,
  \eql{sum7}$$
leading to the generalisation of (\peq{sum5}), for example,
  $$\eqalign{ \sum_{r=[\mu]+1}^\infty{(-1)^r\al^{r+\la}\over r+\la}\,
  \ze(r,w)= \int_0^\al dv&
  v^\la\big(\psi^{(1)}(w+v)-\psi^{(1)}(w)\big)\,\cr
  &-\sum_{r=2}^{[\mu]}{\al^{r+\la}\over (r+\la)(r-1)!}\,
  \psi^{(r)}(w)\,.}  \eql{sum8}$$

One can also easily produce the formula analogous to (\peq{psimom1}) for
the general \zf, (\peq{genzet}),
  $$\eqalign{
  {1\over n!}\int_0^\al dw\, w^n\psi^{(1)}(w)=&-{(-1)^n\over n!}
  \big(\widetilde\ze'(-n)+H_n\,
  \widetilde\ze(-n)\big)\cr
  &+\sum_{l=0}^n{\al^{n-l}\over(n-l)!}{(-1)^l\over l!}
  \big(\ze'(-l,\al)+H_l\,\ze(-l,\al)\big)\,.}
  \eql{psimom2}$$

From the above results one can conclude that the combination
$\ze'(-n,\al)+H_n\,\ze(-n,\al)$ is significant. The best way of
inverting (\peq{psimom2}) for this quantity is to compute the multiple
integral
  $$\eqalign{
  {(-1)^n\over n!}\int_0^\al dw\,(\al- w)^n\psi^{(1)}(w)={(-1)^n\over n!}
  &\big(\ze'(-n,\al)+H_n\,\ze(-n,\al)\big)\cr
  -&\sum_{l=0}^n{(-\al)^{n-l}\over(n-l)!}{(-1)^l\over l!}
  \big(\widetilde\ze'(-l)+H_l\,\widetilde\ze(-l)\big)\,,}
  \eql{riesz}$$
where we have adopted the convention\footnote{Adamchik [\pref{Adamchik1}]
appears to do the same thing in his Proposition 3.} that
  $$
\int_0^\al dw\,\psi^{(1)}(w)=\log\Ga(\al)
  $$
so that (\peq{psimom1}) and (\peq{psimom2}) hold for $n=0$. We confirm this
for the Barnes case, equn. (\peq{psimom1}), the right-hand side of which
is, at $n=0$,
  $$
 \ze'_d(0,\al\mid{\bf d}) - \sum_{i=1}^d \ze'_{d+1-i}(0,d_i\mid{\bf d}_i)
  \,.
  $$

Using the definitions (\peq{defs}), this can be written
  $$\eqalign{
&-\log\bigg( {\rho_d({\bf d}_1)\over
   \Ga_d(\al\mid{\bf d}_1)}.  {\Ga_d(d_1\mid{\bf d}_1)\over\rho_d({\bf d}_1)}.
   {\Ga_{d-1}(d_2\mid{\bf d}_2)\over\rho_{d-1}({\bf d}_2)}\ldots
   {\Ga_1(d_d\mid{\bf d}_d)\over\rho_1({\bf d}_d)}\bigg)\cr
   &=\log\Ga_d(\al\mid{\bf d}_1)}
  $$
where we have used the fact that $\Ga_d(d_*\mid{\bf d})=\rho_{d-1}({\bf
d_*})$ and $\Ga_1(d_d\mid{\bf d}_d)=1$. This is really a check of algebraic
accuracy only, since these relations follow from the recursion
(\peq{iter}).

\section{\bf7. Regularised products and sums.}
With (\peq{sum6}), or (\peq{sum7}), contact has been made with the notion
of regularised products and sums (see \eg [\pref{Voros,QHS,JandL}]) because
the left-hand side is nothing other than the Weierstrass regularisation,
$\sumstar{}\log\big(1+\al/(\la_m+w)\big)$, of the eigenvalue sum
$\sum\log\big(1+\al/(\la_m+w)\big)$,
  $$\eqalign{ \sum_{r=[\mu]+1}^\infty{(-\al)^r\over r}\,\ze(r,w)&=
  -\sumstar{m}\log\bigg(1+{\al\over\la_m+w}\bigg)\cr
  &\equiv-\sum_m\bigg(\log\bigg(1+{\al\over\la_m+w}\bigg)+P\big(-\al/(\la_m+w)
  \big)\bigg)\,,} \eql{regsum}$$
with
  $$
  P(x)=x+{x^2\over2}+\ldots+{x^{[\mu]}\over[\mu]}\,.
  $$
This amounts to the subtraction of the first $[\mu]$ terms in the
$\al$--Taylor expansion of $\log\big(1+\al/(\la_m+w)\big)$.

The Barnes equation, (\peq{sum3}), is a {\it generalised} canonical product
expression for the multiple $\Ga$--function.

Equation (\peq{spderiv}) is, of course, an example of (\peq{sum6}) with
$\la_m$ the eigenvalues of a Laplace-type operator. In fact,
differentiation of the $\al$--Taylor series of (\peq{genzet}), with $w\to
w+\al$, constitutes another approach to the derivation of (\peq{sum6}) (\cf
[\pref{Wilton}] eqn.(9) for the Hurwitz function). In the general case we
may refer to (\peq{sum6}) as {\it Voros' relation}, [\pref{Voros}], eqn
(4.12).

Equation (\peq{sum6}) can be obtained by differentiation via a slightly
different route, [\pref{Dow8,Dow12}]. First, one defines the {\it
Weierstrass $\al$--regularised sum},
  $$\eqalign{
\ze^*(s,w,\al)&\equiv\sumstar{m}{1\over(\la_m+w+\al)^s}\cr &\equiv
\sum_m\bigg({1\over(\la_m+w+\al)^s}-{1\over(\la_m+w)^s}
-\sum_{k=1}^M\comb{-s}k{\al^k\over(\la_m+w)^{k+s}}\bigg),}
  \eql{weier}$$
where sufficient terms in the $\al$--Taylor series have been removed in
order to ensure convergence for any selected value of $s$. The integer $M$
determines how many terms are to be subtracted. In terms of the order,
$\mu$, this means that $M=[\mu-s]$.\footnote{ This construction is slightly
different from that used by Quine {\it et al}
 [\pref{QHS}] in an equivalent analysis. They set $M=[\mu]$ for all $s$.
This is just sufficient to encompass $s=0$. Our definition is more
flexible, as we will see, and is in keeping with the work of Dikii
[\pref{Dikii}] and Watson [\pref{Watson2}].} Differentiation with respect
to $s$, and setting $s$ to zero, produces\footnote{ It is therefore
consistent to write ${\ze^*}'={\ze'}^*$. Note also that $\ze^*(s,w,\al)\ne
\ze^*(s,\al,w)$.}
  $$
\ze^{*'}(0,w,\al) =-\sumstar{m}\ln\big(1+\al/(\la_m+w)\big)\,.
\eql{weierl}$$

Further, the summation over $\la_m$ in $\ze^{*}(s,w+\al)$, (\peq{weier}),
can be performed to give the continuation
  $$
\ze^*(s,w,\al)=\ze(s,w+\al)-\ze(s,w)-\sum_{k=1}^M\comb{-s}k
\al^k\ze(s+k,w).
  \eql{regsum2}$$
In particular
  $$
\ze^{*'}(0,w,\al)=\ze'(0,w+\al)-\ze'(0,w) -{\pa\over\pa
s}\sum_{k=1}^{[\mu]}\comb{-s}k\al^k\ze(s+k,w)\bigg|_{s=0}\,,
  \eql{regsum0}$$
which is just (\peq{sum6}), after evaluation of the final term.

This slicker derivation avoids the use of the heat-kernel form of the \zf,
which really acted only as an intermediary. The present approach can be
thought of as the Mellin transform of the previous one.

A similar technique, in a particular case, is used by Nash and O'Connor
[\pref{NandOC}] App.A.

\section{\bf8. Sphere determinants again.}

The general sum in (\peq{sum6}) was motivated by the example (\peq{sum2}),
needed in the computation of a specific (sphere) determinant. The use made
of it depends on how much is known about the spectrum. For the direct
computation of the sphere determinants, obtained by combining
(\peq{spderiv}), (\peq{sum1}), (\peq{sum2}) and (\peq{sum4}), the
expression in (\peq{sum4}) is calculationally explicit since the relevant
quantities can be found from the properties of the Barnes \zf. In
[\pref{Dow1,Dow3}] the final expressions were given in terms of the
derivatives of the Riemann \zf\ at negative integers.\footnote{ The
calculation uses the aforementioned reduction of the Barnes function for
unit degrees into Hurwitz functions. However, what constitutes a `final'
formula is arguable since the Barnes form could justifiably be regarded as
the end, seeing that what remains is `merely' a numerical or cosmetic
affair. For the full--sphere, there is the minor point that one has to
compute both the Dirichlet and Neumann hemisphere determinants. } In the
sphere combination, the nonlocal Barnes remainders, $R_r$, cancel, in
contrast to the computation of Casimir energies, for example.

The further analysis of the hemisphere derivative, (\peq{spderiv}), can be
carried out less specifically by starting from
  $$
  Z(s,w,\al)=\sum_m{1\over\big((\la_m+w)^2-\al^2\big)^s}
  \eql{zeta3}$$
instead of from (\peq{zeta1}). Then
 $$\eqalign{
Z'(0,w,\al)=2\ze'(0,w)+\sum_{r=1}^{[\mu/2]}{\al^{2r} \over
r}&\left(R_{2r}(w)+ {1\over2}N_{2r}(w)H_{r-1}\right)\cr
+&\sum_{r=[\mu/2]+1}^\infty{\al^{2r}\over r}\ze(2r,w)\,,}
  \eql{deriv}$$
where $\ze(s,w)$ is the general \zf, (\peq{genzet}).

The final sum in (\peq{deriv}) follows from the sum in (\peq{sum6}) by
averaging,
  $$\eqalign{
  \sum_{r=[\mu/2]+1}^\infty {\al^{2r}\over r}\,\ze(2r,w)=
  &\log{\Ga(w+\al)\,\Ga(w-\al)\over\Ga^2(w)}\cr
  &-\sum_{r=1}^{[\mu/2]}
  {\al^{2r}\over r}\,\big(R_{2r}(w)+H_{2r-1}N_{2r}(w)\big)\,,}
  \eql{sum10}$$
and the combination with (\peq{deriv}) yields the mentioned cancellation of
the $\ze(s,w)$ remainders, and also of the $\log\Ga^2(w)$, leaving the
formal, but definite, expression
  $$
  Z'(0,w,\al)=\log{\Ga(w+\al)\,\Ga(w-\al)\over\rho^2}
  -\sum_{r=1}^{[\mu/2]}
  {\al^{2r}\over r}H^O_{r-1}N_{2r}(w)
  \eql{gendet}$$
where $H^O_r$ is the odd harmonic number, $H^O_r=\sum_{k=0}^r1/(2k+1)$.

For the $d$-hemisphere, this result is given in [\pref{Dow1}] and, as
mentioned there, it illustrates the fact that the determinant of a product
is not the product of the determinants, at least not if the determinant is
{\it defined} by \zf\ regularisation. The eigenvalues in (\peq{zeta3})
factorise,
  $$
  (\la_m+w)^2-\al^2=\big(\la_m+w-\al\big)\,
  \big(\la_m+w+\al\big)
  \eql{factor}$$
and the first term on the right-hand side of (\peq{gendet}) gives the
product of the determinants of the individual factors. The remainder is a
correction or `anomaly' which was first noticed in physical contexts by
Allen [\pref{Allen}] and by Chodos and Myers [\pref{Chodos2}] and has
attracted more recent mathematical, and physical, attention. It is
trivially zero when $\al=0$.

For the hemisphere, $\la_m={\bf m.d}$ and  $\ze(s,w)$ is the Barnes \zf\
with unit degrees. If the Neumann and Dirichlet hemisphere expressions are
added, so as to give the full--sphere result, the anomaly contributions
cancel in odd dimensions, as can be specifically checked. Actually, for
odd-dimensional closed manifolds, this vanishing is a general result and
follows from properties of the heat--kernel expansion coefficients.

Everything is quite definite in (\peq{gendet}), which could be taken as the
final answer. Expressions for the hemisphere Laplacian determinants
($\al=(d-1)/2$) were given in [{\pref{Dow1}] in terms of the Barnes
function. However, numerical calculation might require one to go further
and express everything in terms of the Hurwitz or Riemann \zf. (This is
arguable.) The details are in [\pref{Dow3}] where a concrete formula is
produced,
  $$\eqalign{
  \ze'_{S^d}(0)={1\over(d-1)!}&\sum_{k=0}^{d-1}\big(1-(-1)^{d+k}\big)\big(
  S^{(k)}_{d-1}+S^{(k+1)}_d\big)\,\ze_R'(-k)\cr
  &-\sum_{r=1}^u{(d-1)^{2r}\over2^{2r}\,r}\,H_{r-1}^O\,N^R_{2r}(d)
+\log(d-1)\,.}
  \eql{sphderiv}$$
Here, $S^{(k)}_j$ are Stirling numbers and $N^R$ is the sum of the Dirichlet
and Neumann hemisphere \zf\ residues,
  $$
  N^R_{2r}(d)=N^N_{2r}(d)+N^D_{2r}(d)\,.
  $$

These residues are given in terms of generalised Bernoulli polynomials for
which there exists the handy calculational form,
  $$
  N^R_{2r}(d)={2^{2r-d-1}\over(d-1)!(2r-2)!}{d^{2r-2}\over dx^{2r-2}}
  \prod^{d-2}_{i=1}(x-i)\bigg|_{x=(d-1)/2}\,,
  $$
and it is easily confirmed that this is zero for odd $d$, although we
actually know this at an earlier stage.

The graph of the results, up to dimension 23, for the Laplacian determinant
shows a curious difference between odd and even dimensions. The values
diverge as $d$ increases. This might not be surprising as there are
fundamental differences between odd and even dimensional spaces, spheres
especially. Other values of $\al$ in (\peq{op}) can be treated in a like
manner.

We now make some comments on related, full-sphere calculations.  Choi and
Srivastava [\pref{ChandS}], following Voros, use (\peq{sum6}) directly for
the two-- and three--sphere. The Laplacian eigenvalues are written in the
usual form, and the standard binomial degeneracies are used. Equation
(\peq{sum6}), with $w=0$, is then taken as an equation for $\log\Ga(\al)$
in terms of $\log\Ga(0)$, which is easily found,
  $$
  \log\Ga(\al)=\log\Ga(0) + \sum_{r=1}^{[\mu]} {(-\al)^r\over
  r}\,\big(R_r+H_{r-1}N_r\big)-\!\sumstar{m}\!\log\bigg(1+{\al\over\la_m}
  \bigg)\,.
  \eql{cands}$$
The correction terms require $R_r$ and $N_r$ to be determined from the same
$\al=0$ information, which is also relatively straightforward. The most
awkward part is the evaluation of the last term, \ie the regularised sum.
This is obtained from the definition, (\peq{regsum}), after using some
specific \zf\ summations. Although quite workable for small dimensions,
this method obscures the general nature of the cancellations that must
occur.

In an interesting paper, Quine and Choi [\pref{QandC}] express the
determinants through regularised sums and produce an explicit formula for
the $d$--sphere using a method comparable to our own, [\pref{Dow3}], and
obtain the equivalents of (\peq{gendet}) and (\peq{sphderiv}). They do not
split the mode problem into Dirichlet and Neumann hemispheres.
 \sgap
 \noindent{\it Bessel approach to the general \zf.}
 \sgap
An alternative continuation for the \zf, (\peq{zeta3}), uses the Bessel
identity mentioned in section 2. This gives
  $$
  Z(s,w,\al)={\sqrt\pi\over\Ga(s)}\int_0^\infty
  d\tau\, K(\tau)\bigg({\tau\over2\al}\bigg)
  ^{s-1/2}\,I_{s-1/2}(\al\tau)\,,
  \eql{bess2}$$
where $K(\tau)$ is the `cylinder' heat-kernel
  $$
  K(\tau)=\sum_m e^{-(\la_m+w)\tau}\,.
  $$

The terminology is a reflection of the fact that the {\it squares},
$(\la_m+w)^2$, are usually the eigenvalues of a second order Laplace-type
operator so that $K(\tau)$ is the heat-kernel of a square-root
(pseudo)-operator, with different locality properties and possible
$\log\tau$ terms in its expansion.

The form (\peq{bess2}) has been exploited by Bytsenko and Williams
[\pref{BandW}] in connection with the multiplicative anomaly.
\section{\bf9. Implicit eigenvalues.}

In the sphere example, the sum, (\peq{sum2}), is simply an intermediate
calculational quantity. It can be made to play a more important role in the
evaluation of determinants especially when the spectrum is not known
explicitly.

The key idea, [\pref{Dow8}], is to turn (\peq{sum7}) around to give a
formula, this time for $\log\Ga(w)$,
  $$
  \log\Ga(w)=\!\sumstar{}\!\log\bigg(1+{\al\over\la_m+w}\bigg)+\log\Ga(w+\al)
  - \sum_{r=1}^{[\mu]} {\al^r\over r!}\,\psi^{(r)}(w)\,,
  \eql{work}$$
and then to observe, trivially, that the right-hand side has to be
independent of $\al$. It thus can be calculated at any convenient value. In
our work [\pref{Dow8,Dow9,Dow12,Dow10}] the natural, infinite mass, limit
$\al\to\infty$, was selected. The $\al$--dependence has to disappear from
the asymptotic form and what remains must be $\log\Ga(w)$. For example the
last term in (\peq{work}) can be disregarded since its $\al$--dependence is
explicit, in finite terms. Note that one is not required to know anything
about the nature of the $\psi^{(r)}$.

The asymptotic behaviour of $\log\Ga(w+\al)$ as $\al\to\infty$ has been
determined ([\pref{Voros, DandC1,Elizalde1}], for
example) in terms of the heat-kernel coefficients, $C_n(w)$, and contains
no $\al$--independent terms, apart from the normalisation, $\log\rho$. Hence
we obtain the equation
  $$
  \Det D^{-1}={\Ga\over\rho}
  \approx\lim_{\al\to\infty}\prodstar{m}\bigg(1+{\al\over\la_m}\bigg)
  \approx\lim_{\al\to\infty}\prod_{m}\bigg(1+{\al\over\la_m}\bigg)
\exp{\sum_m P(-\al/\la_m)}
  \eql{regp}$$
{\it with the understanding} that only the $\al$--independent part of the
right-hand side is to be retained. The notation has been streamlined a
little by dropping explicit reference to the parameter $w$ and taking the
$\la_m$ as the eigenvalues of the generic operator $D$. Keeping only the
$\al$--independent part means that the $P$-polynomial bit of (\peq{regp})
is usually irrelevant.

This equation is particularly valuable when the eigenvalues are given
implicitly as the roots of some equation, $F(\al)=0$, since, under certain
conditions which are often satisfied, the Mittag-Leffler theorem says that
$F(\al)\sim\prod^*_\la(1+\al/\la)$ and the asymptotic behaviour of $F$ is
often a known episode in special function theory, if one is lucky.

In practice, the method is not necessarily straightforward as the function
$F$ often occurs after separation of variables and there are remaining
summations to be dealt with, but it has been applied to $d$--balls for
scalars and higher spin. The method also works nicely for bounded M\"obius
corners [\pref{Dow12}], where, one might note, the sum (\peq{sum3}) occurs
with $\al$ the parameter in the Robin boundary condition.

There are several variants of the above technique but all involve the
asymptotic behaviour of special functions. The spherical cap has been
successfully treated by Barvinsky {\it et al}, [\pref{Barv}], using the
asymptotics of Legendre functions and the ball has been dealt with in
various ways that all require Olver--Bessel asymptotics, \eg
[\pref{BGKE,BKD,kbook}].

\section{\bf10. A Kaluza-Klein interpretation.}
The previous development suggests that we make
$\ze'(-n,\al)+H_n\,\ze(-n,\al)$ the subject of equation (\peq{riesz}) but
rather than the straight derivatives we try to retain a Barnesian
formulation and rewrite this quantity in terms of a new $\Ga$--function.

The best way of doing this is to introduce the \zf\ that yields this
combination as its derivative at 0. This has formal and manipulative
benefits. Therefore define the \zf\ ,
 $$
 \ze^{(n)}(s,w)= {\Ga(s-n)\over\Ga(s)}\,\ze(s-n,w)\,,
 \eql{kkint}$$
so that
  $$
  {\ze^{(n)}}'(0,w)={(-1)^n\over n!}\big(\ze'(-n,w)+H_n\,\ze(-n,w)\big).
  \eql{kkdet}$$
The new $\Ga$--function and modulus, $\Ga^{(n)}(w)$ and $\rho^{(n)}$, are
defined by
  $$
  \log{\Ga^{(n)}(w)\over\rho^{(n)}}= {\ze^{(n)}}'(0,w)\,,
  \eql{ngamma1}$$
and
 $$\eqalign{
 \log\rho^{(n)}&=-\lim_{w\to0}\bigg({\ze^{(n)}}'(0,w)+b_0\log w\bigg)\cr
 \noalign{\vskip5truept}
 &=-\,{{{\widetilde\ze}^{(n)'}}}(0)\cr
 \noalign{\vskip5truept}
 &=-{(-1)^n\over n!}\,\bigg(\widetilde\ze'(-n)+H_n\,\widetilde\ze(-n)\bigg)}
 \eql{nrho}$$
so that $\Ga^{(0)}(w)=\Ga(w)$ and $\rho^{(0)}=\rho$. In (\peq{nrho}), $b_0$
is the number of zero $\la_m$ modes.

Corresponding $\psi$--functions are defined by
  $$
  \psi^{(n,q)}(w)={\pa^q\over\pa w^q}\log\Ga^{(n)}(w)\,,
  \eql{defpsi}$$
to complete the formalism.

Relation (\peq{reln2}) gives the recursion on the additional dimensions,
  $$
{\pa\over\pa w}{\ze^{(n)}}'(0,w)=-{\ze^{(n-1)}}'(0,w)\,.
  \eql{garec}$$

The basic idea is to set $\ze^{(n)}(s,w)$ in place of $\ze(s,w)$ in formal
manipulations and then, if desired, return to $\ze(s,w)$ via (\peq{kkint}).

Equation (\peq{riesz}) now reads in the new notation,
  $$
\log\Ga^{(n)}(\al)={(-1)^n\over n!}\int_0^\al dw\,(\al- w)^n\psi^{(1)}(w)-
\sum_{l=0}^{n-1}{(-\al)^{n-l}\over (n-l)!}\log\rho^{(l)}\,.
  \eql{ngamma2}$$

An interpretation of (\peq{kkint}) is the following. When the $\la_m$ are
the eigenvalues of the Laplacian on a manifold, $\man$, the derivative,
$\widetilde\ze'(0)$, is, up to a factor, the one-loop effective action of
scalar quantum field theory on the `space-time', $\man$. If a mass is added
then one requires $\ze'(0,m^2)$.

In the same situation, the quantity we have denoted by
$\log\big(\Ga^{(n)}(w)/\rho^{(n)}\big)$ is, again up to a factor, the
effective action density on the Kaluza-Klein manifold,
$\oR^{2n}\times\man$. (It is a density in the noncompact factor,
$\oR^{2n}$.) It therefore can be looked upon, when exponentiated, as
determining the functional determinant density on {\it this} manifold. The
Kaluza-Klein eigenvalues are $\la_m+{\bf k}^2+w$ where ${\bf k}$ is a
$2n$-dimensional real vector and the integral over ${\bf k}$ is taken with
a certain $4\pi$ volume normalisation to make (\peq{kkint}) true.

The situation with which we are most concerned is when the $\la_m$ are the
eigenvalues of a linear pseudo-operator on $\man$, as mentioned earlier for
the Barnes \zf. The `Kaluza-Klein' eigenvalues are $\la_m+{\bf k.1}+w$,
where ${\bf k}$ is a real $n$--vector, with $0\le k_i\le\infty$. The
`Kaluza-Klein' manifold, this time, is $\oR^n\times\man$.

In the specific Barnes case, taking $\man$ to be the $d$--torus, the
Kaluza-Klein manifold $\oR^n\times\man$ can be thought of as a
$(d+n)$--torus with $n$ infinite radii.

\section{\bf11. Generalised Glaisher-Kinkelin-Bendersky constants.}

The usual Glaisher-Kinkelin-Bendersky constants, $A_p$, are defined,
[\pref{bend}], as the $N$--independent parts of the sums,
  $$
  \log A_p=\lim_{N\to\infty}\sum_{m=1}^N m^p\log m\bigg|_{N\rm-independent}=
  \lim_{N\to\infty}\log\prod_{m=1}^N m^{m^p}\bigg|_{N\rm-independent}\,,
  \eql{GKB}$$
which, for $p=0$, is Stirling's formula (with the Stirling constant,
$A_0=\sqrt{2\pi}$).

The original, [\pref{Glaisher}], Glaisher constant,
$A_1=\exp\big(1/12-\ze'_R(-1)\big)$, has surfaced occasionally in statistical
physics, \eg [\pref{MandW}], App.B, as noted by Voros, [\pref{Voros}] and the
Bendersky constants in vacuum quantum field theory, \eg [\pref{DandG}].

It is clear from (\peq{GKB}) that $\log A_p$ is related to the Riemann \zf,
$\ze_R'(-p)$. The exact connection can be obtained by going via the Hurwitz
function, noting that, directly from the definition,
  $$
  \log A_p=\lim_{N\to\infty}\big(\ze'_R(-p,N+1)-
  \ze_R'(-p,1)\big)\bigg|_{N\rm-independent}
  \eql{id}$$
and then using, without thought, the standard, large $w$ expansion of
$\ze_R(s,w)$, [\pref{Erdelyi}], to arrive at
  $$
  \log A_p=-\ze_R'(-p)-H_p\,\ze_R(-p)\,.
  \eql{Adam}$$
This simple formula is given by Adamchik [\pref{Adamchik1}].

In our notation,  and normalisation, (see (\peq{kkdet})),
  $$
  \log A_p=-p\,!(-1)^p\,\ze^{(p)'}_R(0)\,,
  $$
and, in this way, one can give the usual Glaisher--Kinkelin--Bendersky
constants a massless Kaluza--Klein {\it determinantal} interpretation.

The $\Ga$--modular forms, $\rho^{(n)}$, (\peq{nrho}), are `new' constants,
and {\it generalised} Glaisher--Kinkelin--Bendersky numbers,
 $$G_n=\rho^{(n)}\,,
 $$
can be defined in the general situation to mimic the result (\peq{Adam}),
which holds for the Hurwitz \zf\ case (when $\widetilde\ze$ is the Riemann
\zf, $\ze_R$ ).

In terms of the $\Ga$--function, the modular form could be defined as the
$w$--independent part of $\log\Ga^{(n)}(w)$ as $w\to\infty$, so that our
new constants are,
  $$
  \log G_n=\log\rho^{(n)}=\lim_{w\to\infty}\,\log\Ga^{(n)}(w)
  \bigg|_{w\rm -independent}\,,
  \eql{modlim}$$
illustrating the formal similarity to (\peq{GKB}).

For the Barnes case, explicit forms follow from (\peq{psimom1}),
   $$
  \log G_n^{(d)}({\bf d})=-{(-1)^n\over n!}\sum_{i=1}^d
  \big(\ze'_{d+1-i}(-n,d_i\mid{\bf d}_i)
  +H_n\,\ze_{d+1-i}(-n,d_i\mid{\bf d}_i)\big)\,.
  $$

\section{\bf 12. Kaluza-Klein regularised sums.}
The use of the Kaluza--Klein ${\ze^{(n)}}(s,w)$ in order to facilitate the
construction of the sphere derivative, $\ze'(-n,a,\al\mid{\bf d})$, (see
(\peq{zeta1})), is contained in [\pref{Cook}]. The presentation of this in
[\pref{Dow11}] is adapted here to a treatment of the general \zf,
(\peq{genzet}), using regularised sums.

The combination (\peq{kkint}) can be introduced into equation
(\peq{regsum2}) to give
  $$
\ze^{(n)*}(s,w,\al)=\ze^{(n)}(s,w+\al)-\ze^{(n)}(s,w)-\sum_{k=1}^M\comb{-s}k
\al^k\ze^{(n)}(s+k,w)\,,
  \eql{regsum3}$$
so that
  $$
\ze^{(n)*'}(0,w,\al)=\ze^{(n)'}(0,w+\al)- \ze^{(n)'}(0,w)-{\pa\over\pa
s}\sum_{k=1}^{[\mu]+n}\comb{-s}k \al^k\ze^{(n)}(s+k,w)\bigg|_{s=0}\,.
  \eql{regsum4}$$

  Evaluation of the last term, produces immediately the equation corresponding
to (\peq{sum6}),
  $$\eqalign{
\sum_{r={[\mu]}+n+1}^\infty{(-\al)^r\over r}\,\ze^{(n)}(r,w)&= \log
{\Ga^{(n)}(w+\al)\over\Ga^{(n)}(w)}- \sum_{r=1}^{[\mu]+n} {(-\al)^{r}\over
r}\big(R^{(n)}_r+ H_{r-1}N^{(n)}_{r}\big)\cr &= \log
{\Ga^{(n)}(w+\al)\over\Ga^{(n)}(w)}- \sum_{r=1}^{[\mu]+n} {\al^{r}\over
r!}\,\psi^{(n,r)}(w) \,,}
  \eql{sum12}$$
where the residue and remainder are defined by
  $$
\ze^{(n)}(s+r,w)\rightarrow {N^{(n)}_r(w)\over s}+R^{(n)}_r(w) \quad{\rm
as}\,\,s\rightarrow 0\,.
  \eql{pole3}$$

The equality,
  $$
  \ze^{(n)*'}(0,w,\al)=\sum_{r={[\mu]}+n+1}^\infty{(-\al)^r\over r}
\,\ze^{(n)}(r,w)\,,
  $$
is shown exactly as in the usual case ($n=0$) by summing over $\la_m$ and
integrating over ${\bf k}$ after the $\al$--expansion of
$\log\big(1+\al/({\bf k}^2+\la_m+w)\big)$, or of $\log\big(1+\al/({\bf
k.1}+\la_m+w)\big)$ depending on the interpretation of the eigenvalues.

Equation (\peq{sum12}) extends (\peq{sum7}) and is our final, formal
generalisation.

If $\ze^{(n)}$ is viewed only as an intermediate quantity  one should
return to the basic \zf, $\ze(s,w),\,=\ze^{(0)}(s,w)$, and express the
$N^{(n)}_r$ and $R^{(n)}_r$ in terms of the $N_r$ and $R_r$ in order to
rewrite the last summation in (\peq{sum12}).

From the definition (\peq{kkint}), the poles of $\ze^{(n)}(s,w)$ at $s=r$
in the summation range, $1\le r\le [\mu]+n$, divide into two sets, those
due to $\Ga(s-n)/\Ga(s)$, \ie $r=1,\ldots,n$, and those coming, possibly,
from the \zf, \ie $r=n+1\ldots,[\mu]+n$.

A straightforward calculation produces,
  $$\eqalign{
  \sum_{r=1}^{[\mu]+n}&
{(-\al)^{r}\over r}\big(R^{(n)}_r+ H_{r-1}N^{(n)}_{r}\big)\cr
&=\sum_{r=1}^n {(-\al)^r\over r!}
{(-1)^{n-r}\over(n-r)!}\bigg(\ze'(r-n,w)+H_{n-r}\ze(r-n,w)\bigg)\cr
&+\sum_{r=n+1}^{[\mu]+n}{(-\al)^r (r-n-1)!\over
r!}\big(R_{r-n}(w)+H_{r-n-1}N_{r-n}(w)\big)\,,}
  $$
and so equation (\peq{sum12}) can be written entirely in terms of the
original \zf, $\ze(s,w)$.

After some shift in the summation variables and multiplication by factors,
one finds the relation,
  $$\eqalign{
  \sum_{k=[\mu]+1}^\infty&{(-\al)^k\over k} \comb {n+k}n^{-1}\,\ze(k,w)\cr
  &=\al^{-n}\big(\ze'(-n,w+\al)+H_n\,\ze(-n,w+\al)\big)\cr
  &-\sum_{k=0}^n \al^{-k}\comb nk\bigg(\ze'(-k,w)+H_k\ze(-k,w)\bigg)\cr
&-\sum_{k=1}^{[\mu]}{(-\al)^k\over k}\comb{n+k}n^{-1}\big(R_k(w)
+H_{k-1}N_k(w)\big)\,,}
  \eql{sum13}$$
which can of course be proved directly without difficulty. One way is given
in [\pref{KKY}], p.13, for the simpler Hurwitz \zf\ ($\mu=1$).
\section{\bf 13. Conclusion and comments.}

In this work, using the Barnes \zf, we have presented improved and
generalised expressions related to recent, and not so recent, work
concerning sums of \zfs\ which arise incidentally in the computation of
sphere determinants.

We have concentrated on the full--sphere results but the power of the Barnes
function shows up when computing quantities on the orbifold factors,
S$^d/\Ga$, where $\Ga$ is a polytope symmetry group. We leave these
considerations, as well as other factorings such as lens spaces, for
another time.

Extensions of our results to forms and spinors, \cf [\pref{DandKi}], is
straightforward and has topological and possible M--theory applications,
[\pref{FandW}].

A determinant interpretation of the Glaisher--Kinkelin--Bendersky constants
has been given, based on the Adamchik form, (\peq{Adam}). These have been
generalised to the Barnes \zf\ case and also to the more general \zf,
(\peq{genzet}).

\newpage
\section{\bf References}
\vskip 5truept
\begin{putreferences}
  \ref{AandD}{Apps,J.S. and Dowker,J.S. \cqg{15}{1998}{1121}.}
  \ref{Barnesa}{Barnes,E.W.: On the Theory of the
  multiple Gamma function {\it Trans. Camb. Phil. Soc.} {\bf 19} (1903) 374.}
  \ref{Barnesb}{Barnes,E.W.: On the asymptotic expansion of integral
  functions of multiple linear sequence, {\it Trans. Camb. Phil. Soc.} {\bf
  19} (1903) 426.}
  \ref{Barv}{A.O.Barvinsky, Yu.A.Kamenshchik and
  I.P.Karmazin \aop {219}{1992}{201}.}
  \ref{BEK}{Bordag,M., E.Elizalde and K.Kirsten: { Heat kernel coefficients
  of the Laplace operator on the D-dimensional ball}, \jmp{37}{1996}{895}.}
  \ref{BGKE}{M.Bordag, B.Geyer, K.Kirsten and E.Elizalde,
\cmp{179}{1996}{215}.}
  \ref{BKD}{M.Bordag, K.Kirsten,K. and Dowker,J.S.\cmp{182}{1996}{371}.}
  \ref{Branson}{Branson,T.P.: Conformally covariant
  equations on differential forms \cpde{7}{1982}{393-431}.}
  \ref{BandG2}{Branson,T.P. and Gilkey,P.B. {\it Comm. Partial Diff. Eqns.}
  {\bf 15} (1990) 245.}
  \ref{BandG}{Branson,T.P. and Gilkey,P.B. \tams{344}{1994}{479}.}
  \ref{BGV}{Branson,T.P., P.B.Gilkey and
  D.V.Vassilevich {\it The Asymptotics of the Laplacian on a manifold with
  boundary} II, hep-th/9504029.}
  \ref{BO}{T.P.Branson and B.\O rsted \pams{113}{1991}{669}.}
  \ref{ChandD}{Peter Chang and J.S.Dowker, \np{395}{1993}{407}.}
  \ref{cheeg1}{Cheeger, J.: Spectral Geometry of Singular
  Riemannian Spaces. \jdg {18}{1983}{575}.}
  \ref{Cook}{A.Cook 1996 {\it PhD
  Thesis}, University of Manchester.}
  \ref{cheeg3}{Cheeger,J.:Analytic torsion and the heat equation. \aom{109}
  {1979}{259-322}.}
  \ref{DandE}{D'Eath,P.D. and G.V.M.Esposito: Local boundary
  conditions for Dirac operator and one-loop quantum cosmology
  \prD{43}{1991}{3234}.}
  \ref{D}{Dowker,J.S. \cqg{11}{1994}{557}.}
  \ref{Dow8}{J.S.Dowker \cqg{13}{1996}{585-610}.}
  \ref{Dow9}{J.S.Dowker {\it Oddball determinants}, hep-th/9507096.}
  \ref{Dowkc}{J.S.Dowker \cqg{11}{1994}{L7-10}.}
  \ref{Dow10}{J.S.Dowker \pl{B366}{1996}{1075-1086}.}
  \ref{Dow11}{J.S.Dowker \jmp{42}{2001}{1501-1532}.}
  \ref{DandA}{Dowker,J.S. and Apps,J.A. \cqg{12}{1995}{1363}.}
  \ref{D2}{Dowker,J.S. \jmp{35}{1994}{4989}; 1995 (Feb.) erratum.}
  \ref{DandA2}{Dowker,J.S. and J.S.Apps, {\it Functional determinants on
  certain domains}. To appear in the Proceedings of the 6th Moscow Quantum
  Gravity Seminar held in Moscow, June 1995; hep-th/9506204.}
  \ref{DABK}{Dowker,J.S., Apps,J.S., Bordag,M. and Kirsten,K.: Spectral
  invariants for the Dirac equation with various boundary conditions
  \cqg{13}{96}{2911-2920}.}
  \ref{EandR}{E.Elizalde and A.Romeo : An integral involving the generalized
  zeta function, {\it International J. of Math. and Phys.} {\bf13} (1994)
  453.}
  \ref{ELV2}{Elizalde, E., Lygren, M. and Vassilevich, D.V. : Zeta
  function for the laplace operator acting on forms in a ball with gauge
  boundary conditions. hep-th/9605026.}
  \ref{ELV1}{Elizalde, E., Lygren, M.
  and Vassilevich, D.V. : Antisymmetric tensor fields on spheres: functional
  determinants and non-local counterterms, \jmp{}{1996}{} to appear. hep-th/
  9602113.}
  \ref{Erdelyi}{A.Erdelyi,W.Magnus,F.Oberhettinger and
  F.G.Tricomi {\it Higher Transcendental Functions} Vol.I McGraw-Hill, New
  York, 1953.}
  \ref{Gilkey1}{Gilkey, P.B, Invariance theory, the
  heat equation and the Atiyah-Singer index theorem, 2nd. Edn., CRC Press,
  Boca Raton 1995.}
  \ref{IandT}{Ikeda,A. and Taniguchi,Y.:Spectra and
  eigenforms of the Laplacian on $S^n$ and $P^n(C)$, \ojm{15}{1978}{515-546}.}
  \ref{IandK}{Iwasaki,I. and Katase,K. :On the
  spectra of Laplace operator on $\La^*(S^n)$ \pja{55}{1979}{141}.}
  \ref{KandC}{Kirsten,K. and Cognola.G,: {
  Heat-kernel coefficients and functional determinants for higher spin fields
  on the ball} \cqg{13}{1996} {633-644}.}
  \ref{KKY}{S.Kanemitsu, M.Katsurada and M.Yoshimoto {\it Aeq. Math.}
  {\bf 59} (2000) 1-19.}
  \ref{NandOC}{C.Nash and D-J.O'Connor \jmp{36}{1995}{1462}.}
  \ref{norlund2}{N\"orlund~N. E.:M\'emoire sur les
  polynomes de Bernoulli. \am {4}{1921} {1922}.}
  \ref{RandT}{Russell,I.H. and Toms D.J.: Vacuum energy
  for massive forms in $R^m\times S^N$, \cqg{4}{1986}{1357}.}
  \ref{Rayleigh}{Rayleigh,Lord, {\it The Theory of Sound}, Vols. I and II,
  2nd Edn., (MacMillan, London, 1894).}
  \ref{Shint}{T.Shintani {\it J.Fac.Tokyo} {\bf 24} (1977) 167-199.}
  \ref{Vign}{M-F.Vigneras {\it Ast\'erisque} {\bf 61} (1979) 235-249.}
  \ref{Voros}{Voros,A.: Spectral
  functions, special functions and the Selberg zeta function,
\cmp{110}{1987}439-465.}
  \ref{Weis}{W.I.Weisberger \cmp {112}{1987}{633}.}
  \ref{Ray}{Ray,D.B.: Reidemeister torsion and the Laplacian on lens spaces
  \aim{4}{1970}{109}.}
  \ref{McandO}{McAvity,D.M. and Osborn,H. Asymptotic
  expansion of the heat kernel for generalised boundary conditions,
  \cqg{8}{1991}{1445}.}
  \ref{barv}{Barvinsky,A.O.\pl{195B}{1987}{344}.}
  \ref{BGKE}{M.Bordag, B.Geyer,
  K.Kirsten and E.Elizalde, \cmp{179}{1996}{215}.}
  \ref{CandD}{Peter Chang and J.S.Dowker \np {395}{1993}{407}.}
  \ref{Allen}{B.Allen, PhD Thesis, University of Cambridge, 1984.}
  \ref{Weisbergerb}{W.I.Weisberger \cmp{112}{1987}{633}.}
  \ref{Hortacsu}{M.Hortacsu, K.D.Rothe and B.Schroer \prD
   {20}{1979}{3203-3212}.}
  \ref{Weisbergera}{W.I.Weisberger \np{284}{1987}{171}.}
  \ref{Chodos1}{A.Chodos and E.Myers \aop{156}{1984}{412}.}
  \ref{Chodos2}{A.Chodos and E.Myers {\it Can.J.Phys.} {\bf 64} (1986) 633.}
  \ref{QHS}{J.R.Quine, S.H.Heydari and R.Y.Song \tams{338}{1993}{213}.}
  \ref{Minak}{S.Minakshisundaram {\it J. Ind. Math. Soc.} {\bf 13} (1949) 41.}
  \ref{Camporesi}{R.Camporesi {\it Physics Reports} {\bf 196} (1990) 1.}
  \ref{CandW}{P.Candelas and S.Weinberg \np{237}{1984}{397}.}
  \ref{Vardi}{I.Vardi, {\it SIAM J. Math. Anal.} {\bf 19} (1988) 493-507.}
  \ref{Gillet}{H.Gillet, C.Soul\'e and D.Zagier \top{30}{1992}{21}.}
  \ref{Aurell}{E.Aurell and P.Salomonson {\it On functional determinants of
  Laplacians in polygons and simplices}}
   \ref{Dowkb}{J.S.Dowker \pr{36}{1986}{1111}.}
   \ref{Barv}{A.O.Barvinsky, Yu.A.Kamenshchik and I.P.Karmazin \anp {219}
   {1992}{201}.}
   \ref{Kam}{Yu.A.Kamenshchik and I.V.Mishakov {\it Int. J. Mod. Phys.}
  {\bf A7} (1992) 3265.}
  \ref{Rayleigh}{Lord Rayleigh{\it Theory of Sound} vols.I and II,
  MacMillan, London, 1877,78.}
  \ref{KCD}{G.Kennedy, R.Critchley and J.S.Dowker \aop{125}{1980}{346}.}
  \ref{Donnelly}{H.Donnelly \ma{224}{1976}{161}.}
  \ref{Fur2}{D.V.Fursaev {\sl Spectral geometry and one-loop divergences on
  manifolds with conical singularities}, JINR preprint DSF-13/94,
  hep-th/9405143.}
  \ref{HandE}{S.W.Hawking and G.F.R.Ellis {\sl The large scale structure of
  space-time} Cambridge University Press, 1973.}
  \ref{DandK}{J.S.Dowker and G.Kennedy \jpa{11}{1978}{895}.}
  \ref{ChandD}{Peter Chang and J.S.Dowker \np{395}{1993}{407}.}
  \ref{FandM}{D.V.Fursaev and G.Miele \pr{D49}{1994}{987}.}
  \ref{Dowkerccs}{J.S.Dowker \cqg{4}{1987}{L157}.}
  \ref{BandH}{J.Br\"uning and E.Heintze \dmj{51}{1984}{959}.}
  \ref{Cheeger}{J.Cheeger \jdg{18}{1983}{575}.}
  \ref{SandW}{K.Stewartson and R.T.Waechter \pcps{69}{1971}{353}.}
  \ref{CandJ}{H.S.Carslaw and J.C.Jaeger {\it The conduction of heat
  in solids} Oxford, The Clarendon Press, 1959.}
  \ref{BandH}{H.P.Baltes and E.M.Hilf {\it Spectra of finite systems}.}
  \ref{Epstein}{P.Epstein \ma{56}{1903}{615}.}
  \ref{Kennedy2}{G.Kennedy PhD thesis, Manchester (1978).}
  \ref{Kennedy3}{G.Kennedy \jpa{11}{1978}{L173}.}
  \ref{Luscher}{M.L\"uscher, K.Symanzik and P.Weiss \np {173}{1980}{365}.}
  \ref{Polyakov}{A.M.Polyakov \pl {103}{1981}{207}.}
  \ref{Bukhb}{L.Bukhbinder, V.P.Gusynin and P.I.Fomin {\it Sov. J. Nucl.
  Phys.} {\bf 44} (1986) 534.}
  \ref{Alvarez}{O.Alvarez \np {216}{1983}{125}.}
  \ref{DandS}{J.S.Dowker and J.P.Schofield \jmp{31}{1990}{808}.}
  \ref{Dow1}{J.S.Dowker \cmp{162}{1994}{633}.}
  \ref{Dow2}{J.S.Dowker \cqg{11}{1994}{557}.}
  \ref{Dow3}{J.S.Dowker \jmp{35}{1994}{4989}; erratum {\it ibid}, Feb.1995.}
  \ref{Dow4}{J.S.Dowker \pr{D36}{1987}{3095}}
  \ref{Dow5}{J.S.Dowker {\it Heat-kernels and polytopes}}
  \ref{Dow6}{J.S.Dowker \pr{D50}{1994}{6369}.}
  \ref{Dow7}{J.S.Dowker \pr{D39}{1989}{1235}.}
  \ref{Dow13}{J.S.Dowker \cqg{1}{1984}{359}.}
  \ref{BandG}{P.B.Gilkey and T.P.Branson \tams{344}{1994}{479}.}
  \ref{Schofield}{J.P.Schofield Ph.D.thesis, University of Manchester,
  (1991).}
  \ref{Barnesa}{E.W.Barnes {\it Trans. Camb. Phil. Soc.} {\bf 19} (1903)
  374.}
  \ref{Barnesb}{E.W.Barnes {\it Trans. Camb. Phil. Soc.} {\bf 19} (1903)
  426.}
  \ref{BandG2}{T.P.Branson and P.B.Gilkey {\it Comm. Partial Diff. Equations}
  {\bf 15} (1990) 245.}
  \ref{Pathria}{R.K.Pathria {\it Suppl.Nuovo Cim.} {\bf 4} (1966) 276.}
  \ref{Baltes}{H.P.Baltes \prA{6}{1972}{2252}.}
  \ref{Moss}{I.Moss \cqg{6}{1989}{659}.}
  \ref{Barv}{A.O.Barvinsky, Yu.A.Kamenshchik and I.P.Karmazin \aop {219}
  {1992}{201}.}
  \ref{Kam}{Yu.A.Kamenshchik and I.V.Mishakov \prD{47}{1993}{1380}.}
  \ref{KandM}{Yu.A.Kamenshchik and I.V.Mishakov {\it Int. J. Mod. Phys.}
  {\bf A7} (1992) 3265.}
  \ref{DandE}{P.D.D'Eath and G.V.M.Esposito \prD{43}{1991}{3234}.}
  \ref{Rich}{K.Richardson \jfa{122}{1994}{52}.}
  \ref{Osgood}{B.Osgood, R.Phillips and P.Sarnak \jfa{80}{1988}{148}.}
  \ref{BCY}{T.P.Branson, S.-Y. A.Chang and P.C.Yang \cmp{149}{1992}{241}.}
  \ref{Vass}{D.V.Vassilevich.{\it Vector fields on a disk with mixed
  boundary conditions} gr-qc /9404052.}
  \ref{MandP}{I.Moss and S.Poletti \pl{B333}{1994}{326}.}
  \ref{Aurell1}{E.Aurell and P.Salomonson \cmp{165}{1994}{233}.}
  \ref{Aurell2}{E.Aurell and P.Salomonson {\it Further results on functional
  determinants of laplacians on simplicial complexes} hep-th/9405140.}
  \ref{BandO}{T.P.Branson and B.\O rsted \pams{113}{1991}{669}.}
  \ref{Elizalde1}{E.Elizalde \jmp{36}{1994}{3308}.}
  \ref{BandK}{M.Bordag and K.Kirsten {\it Heat-kernel coefficients of
  the Laplace operator on the 3-dimensional ball} hep-th/9501064.}
  \ref{Waechter}{R.T.Waechter \pcps{72}{1972}{439}.}
  \ref{CandC}{A.Capelli and A.Costa \np {314}{1989}{707}.}
  \ref{BandH}{M.V.Berry and C.J.Howls \prs {447}{1994}{527}.}
  \ref{DandW}{A.Dettki and A.Wipf \np{377}{1992}{252}.}
  \ref{Weisbergerb} {W.I.Weisberger \cmp{112}{1987}{633}.}
  \ref{Voros}{A.Voros \cmp{110}{1987}{439-465}.}
  \ref{Pockels}{F.Pockels {\it \"Uber die partielle Differentialgleichung
  $\Delta u+k^2u=0$}, B.G.Teubner, Leipzig 1891.}
  \ref{Kober}{H.Kober \mz{39}{1935}{609}.}
  \ref{Watson2}{G.N.Watson \qjm{2}{1931}{300}.}
  \ref{DandC1}{J.S.Dowker and R.Critchley \prD {13}{1976}{3224}.}
  \ref{Lamb}{H.Lamb \pm{15}{1884}{270}.}
  \ref{EandR}{E.Elizalde and A.Romeo {\it International J. of Math. and
  Phys.} {\bf13} (1994) 453.}
  \ref{DandA}{J.S.Dowker and J.S.Apps \cqg{12}{1995}{1363}.}
  \ref{Watson1}{G.N.Watson {\it Theory of Bessel Functions} Cambridge
  University Press, Cambridge, 1944.}
   \ref{MandO}{W.Magnus and F.Oberhettinger {\it Formeln und S\"atze}
  Springer-Verlag, Berlin, 1948.}
  \ref{Olver}{F.W.J.Olver {\it Phil.Trans.Roy.Soc} {\bf A247} (1954) 328.}
  \ref{Hurt}{N.E.Hurt {\it Geometric Quantization in action} Reidel,
  Dordrecht, 1983.}

  \ref{Louko}{J.Louko \prD{38}{1988}{478}.}
  \ref{Schleich} {K.Schleich \prD{32}{85}{1989}.}
  \ref{ELZ}{E.Elizalde, S.Leseduarte and S.Zerbini. hep-th/9303126.}
  \ref{BGV}{T.P.Branson, P.B.Gilkey and D.V.Vassilevich {\it The Asymptotics
  of the Laplacian on a manifold with boundary} II, hep-th/9504029.}
  \ref{Erdelyi}{A.Erdelyi,W.Magnus,F.Oberhettinger and F.G.Tricomi {\it
  Higher Transcendental Functions} Vol.I McGraw-Hill, New York, 1953.}
  \ref{Dikii}{L.A.Dikii {\it Usp. Mat. Nauk.} {\bf13} (1958) 111.}
  \ref{WandW}{E.T.Whittaker and G.M.Watson {\it A Course in Modern Analysis},
  Cambridge University Press, Cambridge, 1927.}
  \ref{JandL}{J.Jorgenson and S.Lang Lect.Notes in Math. 1564 Springer-Verlag,
  Berlin 1993.}
  \ref{ChandS}{J.Choi and H.M.Srivastava {\it Kyushu J.Math.}
{\bf 53} (1999) {209-222}.}
  \ref{CandN}{J.Choi and C.Nash {\it Math. Japon} {\bf 45} (1997) 223-230.}
  \ref{QandC}{J.R.Quine and J.Choi \rmjm {26}{1996}{719-729}.}
  \ref{Choi}{J.Choi {\it Math. Japon.} {\bf 40} (1994) 155-166.}
  \ref{kumagai}{H.Kumagai {\it Acta Arithmetica} {\bf 91} (1999) 199-208.}
  \ref{bend}{L.Bendersky \am{61}{1933}{263-322}.}
  \ref{kink}{(H).Kinkelin J.f.reine u. angew. Math. (Crelle) {\bf 57} (1860)
   122-158.}
  \ref{holder}{O.H\"older {\it G\"ott. Nachrichten} (1886) 514-522.}
  \ref{alex}{W.P.Alexeiewsky {\it Leipzig Berichte} {\bf 46} (1894) 268-275.}
  \ref{Adamchik1}{V.S.Adamchik {\it J. Comp. Appl. Math.} {\bf 100} (1998)
  191-199.}
  \ref{Dow12}{J.S.Dowker, Functional Determinants on M\"obius corners in
  {\it Quantum Field Theory Under the Influence of External Conditions},
  ed. by M.Bordag, Teubner, Leipzig, 1996.}
  \ref{EandM}{O.Espinosa and V.H.Moll, {\it The Ramanujan Journal} {\bf6}
(2002) 449.}
  \ref{BandW}{A.Bytsenko and F.Williams \jmp{39}{1998}{1075-1086}.}
  \ref{DandKi}{J.S.Dowker and Klaus Kirsten {\it Comm. Anal. and Geom.}
  {\bf 7} (1999) 641-679.}
  \ref{KandT}{Klaus Kirsten and D.J.Toms \prA{54}{1996}{4188}.}
  \ref{MandW}{B.McCoy and T.T.Wu {\it The two-dimensional Ising model},
  Harvard University Press, Cambridge, 1973.}
  \ref{Wilton}{J.R.Wilton, \mom{52}{1922/23}{90-93}.}
  \ref{Sarnak}{P.Sarnak, \cmp{110}{1987}{102-109}.}
  \ref{Sriv}{H.M.Srivastava and J.Choi, {\it Kluwer series on Mathematics
  and its
  Applications}, Vol. {\bf 531}, Kluwer, Dordrecht, 2001.}
  \ref{JandM}{M.Jimbo and T.Miwa \jpa{29}{1996}{2923-2958}.}
  \ref{Kuro}{N.Kurokawa, {\it Proc.Jap.Acad} {\bf A67}(1991) 61-64; {\it ibid}
  {\bf A68}(1992) 256-260.}
  \ref{Ruijsenaars}{S.N.M.Ruijsenhaars \aim{156}{2000}{107-132}.}
  \ref{Smirnov}{F.A.Smirnov, Adv. Series in Math. Phys. Vol. {\bf14}, World
  Scientific, Singapore, 1992.}
  \ref{Luk}{S.Lukyanov, \cmp{167}{1995}{183-226}.}
  \ref{Illies}{G.Illies, \cmp{220}{2001}{69-94}.}
  \ref{Glaisher}{J.W.L.Glaisher \mom{6}{1877}{71-76}; {\it ibid} {\bf 7}
  (1878) 43-47; {\it ibid} {\bf 23} (1893)  145-175; {\it ibid} {\bf 24}
  (1894) 1-16.}
  \ref{FandZ}{P.G.O.Freund and A.V.Zabrodin \jmp{34}{1993}{5832}.}
  \ref{KKY}{S.Kanemitsu, H.Kumagai and M.Yoshimoto, {\it The Ramanujan J.}
  {\bf 5} (2001) {5}.}
  \ref{FandW}{T.Friedmann and E.Witten {\it Unification scale, proton decay
  and manifolds of G$_2$ holonomy}, hep-th/0211269.}
  \ref{Manin}{Y.Manin {\it Ast\'erisque} {\bf 228} (1995) 121.}
  \ref{DandG}{W.Dittrich and H.Gies {\it Probing the quantum vacuum}, Springer
Tracts Mod. Phys. {\bf 166} (2000) 1.}
  \ref{kbook}{K.Kirsten {\it Spectral functions in mathematics and physics}
(CRC, Boca Raton, 2001).}
  \ref{HKK}{M.Holthaus, E.Kalinowski and K.Kirsten \aop{270}{1998}{137}.}
\end{putreferences}
\bye